\documentclass[a4paper,twoside]{article}
\usepackage{graphicx}
\usepackage{hyperref}
\usepackage{xspace}
\usepackage{fancybox}
\usepackage{subfig}

\usepackage{epsfig}
\usepackage{calc}
\usepackage{amssymb}
\usepackage{amstext}
\usepackage{amsmath}
\usepackage{amsthm}
\usepackage{multicol}
\usepackage{pslatex}
\usepackage{apalike}
\usepackage{eurosym}
\usepackage{SCITEPRESS}     

\begin{document}

\title{A Cloud-Based Collaboration Platform for Model-Based Design of Cyber-Physical Systems}

\author{\authorname{Peter Gorm Larsen\sup{1}\orcidAuthor{0000-0002-4589-1500},
  Hugo Daniel Macedo\sup{1}\orcidAuthor{0000-0002-8882-4500},
  John Fitzgerald\sup{2}\orcidAuthor{0000-0001-7041-1807},
  Holger Pfeifer\sup{3},
  Martin Benedict\sup{4},
  Stefano Tonetta\sup{5}\orcidAuthor{0000-0001-9091-7899},
  Angelo Marguglio\sup{6},
  Sergio Gusmeroli \sup{7} and
  George Suciu Jr.\sup{8}
  }
\affiliation{\sup{1}DIGIT, Department of Engineering, Aarhus University, Aarhus, Denmark}
\affiliation{\sup{2}School of Computing, Newcastle University, United Kingdom}
\affiliation{\sup{3}fortiss, Germany}
\affiliation{\sup{4}Virtual Vehicle, Austria}
\affiliation{\sup{5}Fondazione Bruno Kessler, Italy}
\affiliation{\sup{6}Engineering Ingegneria Informatica S.p.A., Italy}
\affiliation{\sup{7}Politecnico di Milano, Italy}
\affiliation{\sup{8}BEIA Consult, Romania}
\email{\{pgl,hdm\}@eng.au.dk,
John.Fitzgerald@ncl.ac.uk,
pfeifer@fortiss.org,
martin.benedikt@v2c2.at,
tonettas@fbk.eu,
angelo.marguglio@eng.it,
sergio.gusmeroli@polimi.it,
george@beia.ro}
}

\abstract{
Businesses, particularly small and medium-sized enterprises, aiming to start up in Model-Based Design (MBD) face difficult choices from a wide range of methods, notations and tools before making the significant investments in planning, procurement and training necessary to deploy new approaches successfully. In the development of Cyber-Physical Systems (CPSs) this is exacerbated by the diversity of formalisms covering computation, physical and human processes. In this paper, we propose the use of a cloud-enabled and open collaboration platform that allows businesses to offer models, tools and other assets, and permits others to access these on a pay-per-use basis as a means of lowering barriers to the adoption of MBD technology, and to promote experimentation in a sandbox environment.
}

\onecolumn \maketitle \normalsize \setcounter{footnote}{0} \vfill

\section{\uppercase{Introduction}} \label{sec:intro}

\noindent The digital transformation that industry and society are experiencing creates many opportunities for increasing the delegation of tasks to machines. The dependability of the resulting systems therefore becomes critical not only for compliance with regulations and certification standards, but also for societal acceptance of the transformation. Modern innovative products and systems combine physical and networked computational processes. Such Cyber-Physical Systems (CPSs) place new multi-disciplinary demands on traditional engineering processes because of the heterogeneity and complexity of the constituent elements, and of their interaction with the environment.

Model-Based Design (MBD) has demonstrated the potential to increase the quality of CPSs~\cite{VanderAuweraer&13}. MBD prescribes the use of system models through the development process in order to represent system structure and behaviours, providing a basis for machine-assisted analysis of system properties, and informing design decisions through processes of refinement into implementation.

A considerable body of research has provided model-based solutions to the challenges of CPS design~\cite{Beckers&07,Sztipanovits&15}, but businesses that could benefit from such approaches may face barriers to their adoption. It is possible that as a consequence, MBD methods and tools appear largely to be applied in domains such as aerospace where the return of investment can take decades. By contrast, SMEs require considerable flexibility to change processes to adopt MBD, and may lack in-house expertise. In addition, the selection, procurement, training and deployment costs for some methods and tools can be discouragingly high.  In general, it is difficult for SMEs to invest in acquiring the necessary background for example because of the high license fees from commercial vendors of MBD assets.


In this position paper, we report on an approach that aims to make MBD more accessible to a range of businesses, but especially SMEs, involved in the development of cyber-physical products and systems. This centres on the servitisation of modelling, simulation and analysis tools for CPS design, offered through an open collaboration platform. The new HUBCAP project\footnote{See \url{http://www.hubcap.eu}.} aims to implement and evaluate this approach, alongside provision of other services for SMEs through a network of innovation hubs. In this paper, however, we focus on the platform.

Our goal is to provide a collaboration platform that enables users to access advanced CPS design and engineering solutions, including models and tools such as those offered, for example, by the Modelica association community \cite{Fritzson15} or the INTO-CPS association community~\cite{Larsen&16a}.
This will be done through \emph{sandboxes} -- environments in which users can test and experiment with a solution in a secure and trusted environment before investing in longer-term or larger-scale adoption. The platform will also facilitate collaboration services that enable the sharing of knowledge among providers of MBD solutions, so that new models, tools and techniques and related services may be extended by combining existing assets.

This paper provides an overview of our proposed collaboration platform in Section~\ref{sec:platform}, and more detail on the sandbox functionality in Section~\ref{sec:sandbox}. Section~\ref{sec:models} indicates how assets such as models can be accessed from a catalogue inside the collaboration platform. Afterwards Section~\ref{sec:opencalls} explains how we envisage the collaboration platform to be populated with MDB assets by means of open calls targeted SMEs enabling them to get financial support. Finally, we conclude in Section~\ref{sec:conclude} by discussing the work required to realise and evaluate such a platform.

\section{\uppercase{The Collaboration Platform}}\label{sec:platform}

The HUBCAP Collaboration Platform is based on the DIHIWARE open source solution
developed by the MIDIH H2020 EU project\footnote{See \url{http://midih.eu/}.}  and currently
in use in many ecosystems in Europe \cite{Kainz&19}. DIHIWARE offers a
complete collaboration environment inspired by Enterprise Social Software~\cite{Cook08}. It
supports both ``Access to'' and ``Collaborate with'' services, providing
companies access to the latest knowledge, expertise and technology during their
digital transformation paths toward piloting, testing and experimenting with
new digital technologies.

The knowledge-driven services, complemented by the collaborative and innovation
side of the platform, are intended to create a virtual environment where providers and
consumers of digital technologies are not just matching assets and needs, but
they are collaborating together towards joint innovations. This environment
will be the foundation on which specific customisations (environment
customisation, catalogue designing, and dedicated user journey) will be realised
to meet the specific needs of HUBCAP.

\begin{figure*}
  \centering
  \subfloat[DIHIWARE Platform]{\includegraphics[width=0.49\textwidth]{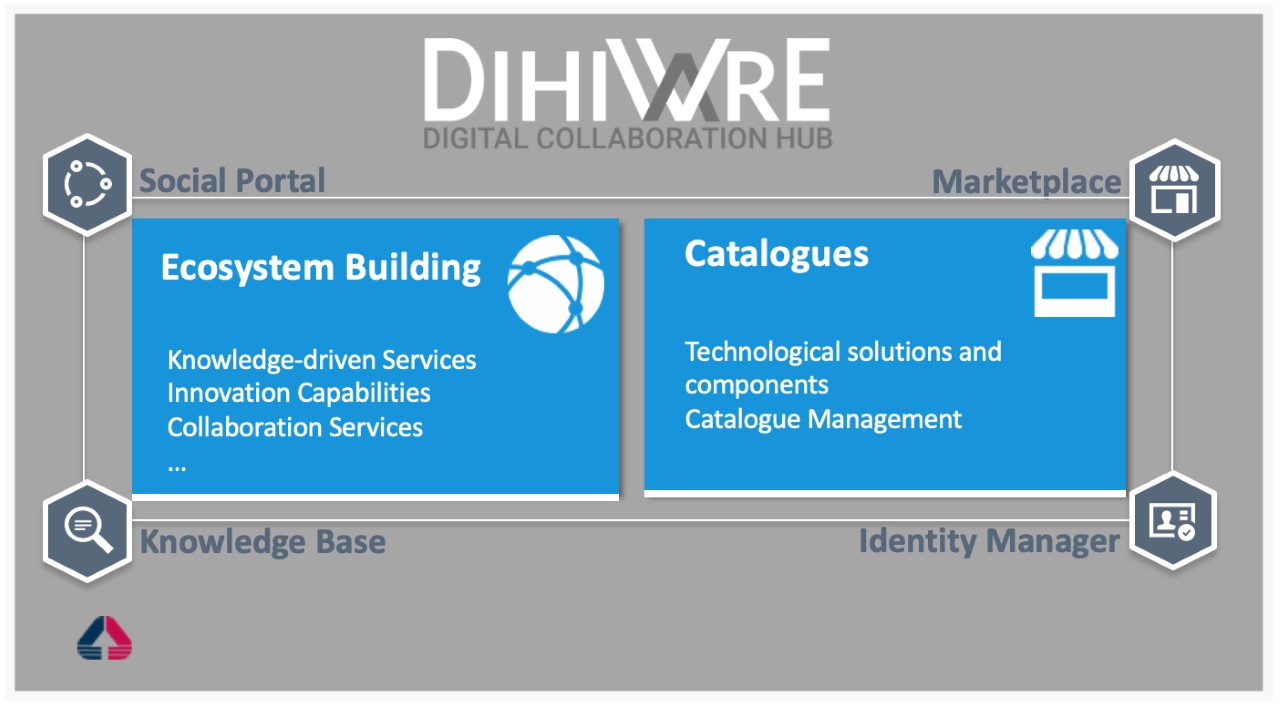}\label{fig:dih_hla}}
  \hspace{0.01\textwidth}\hfill
  \subfloat[The HUBCAP Platform]{\includegraphics[width=0.445\textwidth]{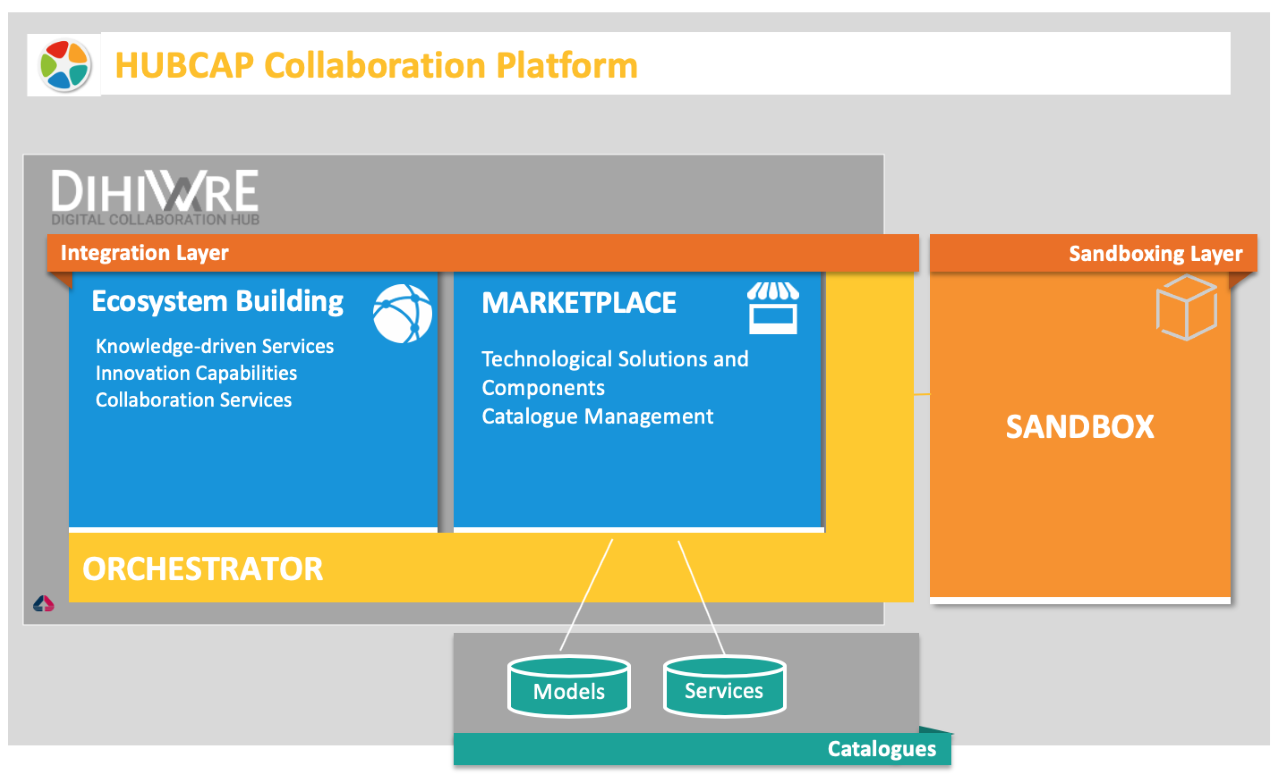}\label{fig:hubcap_platform}}
\caption{Growing from the DIHIWARE to the HUBCAP Platform}
\end{figure*}

The platform integrates open source technologies (e.g., those coming
from the FIWARE Community\footnote{See \url{http://fiware.org/}.} with enterprise-grade solutions
such as Liferay Portal. DIHIWARE has four main subsystems~(Figure~\ref{fig:dih_hla}):

\begin{description}
\item[Identity Manager:] This subsystem centralises user authentication, defining roles and granting access while using the other applications.

\item[Marketplace:] This subsystem handles the creation of company offerings by means of a product catalogue in which MBD assets and services will be shared. End users can view and interact with assets through this subsystem, while suppliers use it to manage their asset and service catalogue.

\item[Knowledge base:] This subsystem supports semantic indexing and retrieval functions, grounded on semantic technologies and providing a set of services for creating awareness, dissemination, training, and managing connections among user-generated content.

\item[Social Portal:] This subsystem offers tools for social activity, user collaboration, matchmaking, and expert search, one of the key offerings on DIHIWARE.
\end{description}

HUBCAP extends this foundation framework with the sandbox capability. The overall idea is that it shall be the centre of an ecosystem where different organisations can collaborate in a model-based fashion.
In order to protect intellectual property of the suppliers of assets, the platform needs to enable white-box, grey-box and black-box models so that it is possible to control access. This can for example be accomplished using the Functional Mockup Interface (FMI) standard that is supported by range of current tools \cite{FMIStandard2.0} to enable co-simulation of a collection of different individual models can be enabled \cite{Gomes&18}.

The platform's collaboration functions will enable diverse partners to work together through the sharing of controlled access to models. We imagine that an Original Equipment Manufacturer (OEM) may invite their suppliers to join the model-based development of a CPS. Many of the suppliers will wish to provide the models as black boxes. For the OEM this is not a problem as long these diverse models can be integrated in their analysis of CPS-level properties. The idea is that all the partners will be able collectively to analyse the composition from the sandbox.

\section{\uppercase{The Sandbox Functionality}}\label{sec:sandbox}
\label{sec:tsf}

In a typical MBD setup there are three classes of assets:
\begin{itemize}
	\item models, which are mathematical or formal abstractions of system elements (components or subsystems);
	\item tools, which are software packages and their dependencies that enable the development, analysis, and simulation of models; and
	\item Operating System (OS), which refers to a software environment providing libraries and dependencies needed to run the tools.
\end{itemize}


Adding a sandbox feature to the DIHIWARE collaboration platform requires the support of a multitude of tools, dependencies, OSs. This calls for a virtualisation approach in which the tools involved in a given experiment are deployed for each individual user in an environment that is populated with all the dependencies required. In addition to plain virtualisation, a sandbox mechanism provides improved security and minimises perceived interference among users. We envisage three kinds of sandboxes, each with a specific complexity and level of service:

\begin{figure*}[bt]\centering
\includegraphics[width=\textwidth]{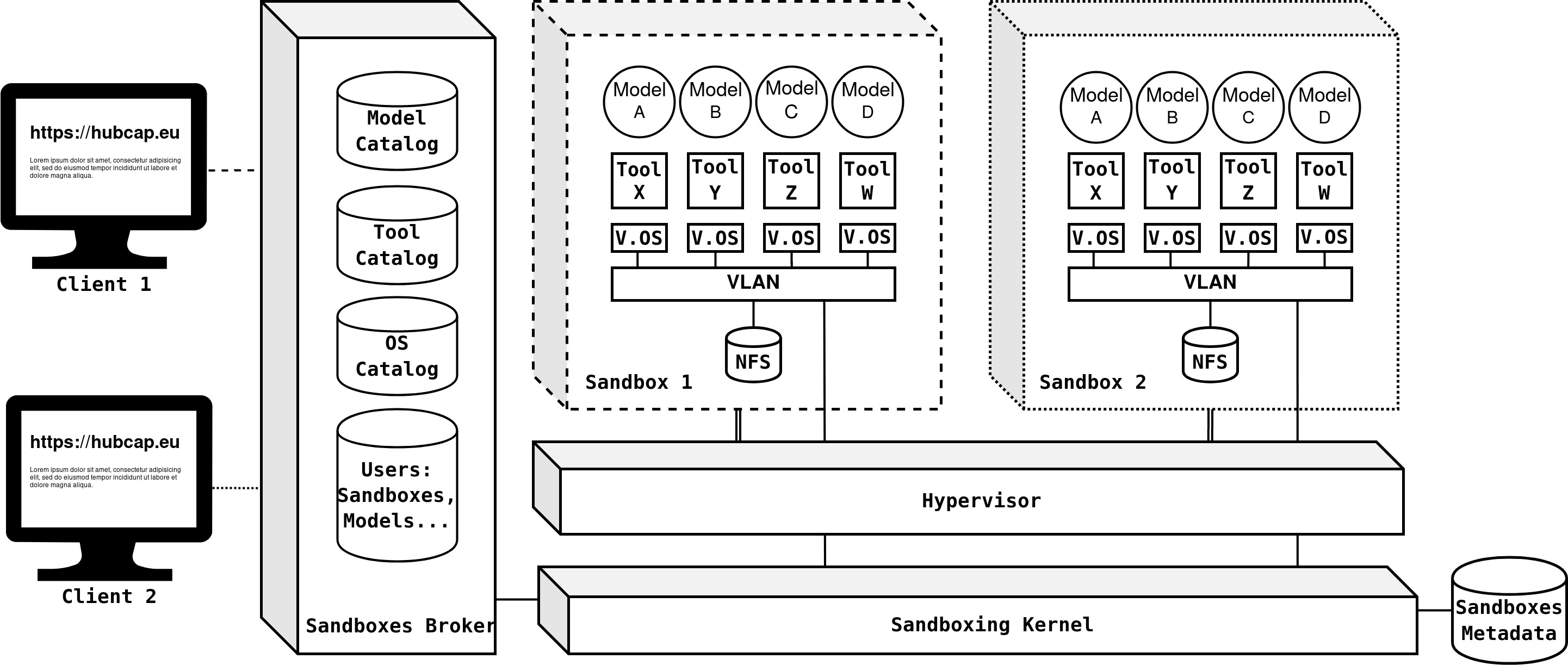}
\caption{\label{fig:sandbox}The HUBCAP Sandbox architecture}
\end{figure*}

\begin{description}
\item[Demo Sandbox:] A simple web application demonstration, where inputs may be predefined for user experimentation. This supports simple cases in which a simple web site or application is sufficient to provide users with the inner workings of an experiment, serving as a potentially advanced/animated repository of the models and shared artefacts, with pointers on to the vendor sites. This serves as a solution for tool vendors/experiments with a need to host results without the need for significant dependence on the underlying OS features. 

\item[Container Sandbox:] A web application deployed using a standard virtualised processes approach in which different containers host different services of the application and are connected through a virtual network interface. This option allows experiments involving tools requiring a single  OS, but needing complex software environments (different versions of an environment) or a multiple services running in different containers.

\item[Cloud Sandbox:] A single sandbox is implemented by a set of virtual machines (VMs) operating on top of a Kernel-based Virtual Machine (KVM), connected through a KVM virtual network. The VMs inside a sandbox can access a common NFS storage to retrieve some input data (e.g., existing models) and to exchange data among themselves (e.g., generated models). This alternative supports experiments which are based on tools requiring different OSs or OS versions are allowed to run, exchange data via the network, and exchange files.
\end{description}

It is possible to realise a Demo or Container Sandbox using the Cloud type. A working prototype of this approach has been constructed, and details of it follow. The prototype allows a pre-registered HUBCAP Platform user to select a combination of models, tools, and OS, pack them in an isolated sandbox and
start playing with them via a web browser. Examples of use cases for the different kinds of sandboxes are given in Section~\ref{sec:mdbservices}.

\subsection{Architecture}

A sandbox is implemented as an isolated set of VMs (each one running a CPS tool) that interact each other sharing a virtual dedicated subnet and a dedicated NFS storage. No interaction is permitted between the VMs belonging to different sandboxes. The sandbox capability integrated with the DIHIWARE Platform is therefore a sort of private cloud service provider
plus the middleware to manage and mediate the access to those cloud services. In addition, as many cloud service providers offer the capability to select a combination of hardware and operating systems, the HUBCAP Platform offers you to select a combination of OS environments, tools, and models to run an experiment using the HUBCAP sandbox feature.

The sandbox service is outlined in Figure~\ref{fig:sandbox}. The DIHIWARE Platform is enhanced with a broker component (labelled as \emph{Sandboxes Broker} in the figure), which hosts a web application mediating the access of different users (\emph{Client 1} and \emph{Client 2}) to the sandboxes they requested (\emph{Sandbox 1} and \emph{Sandbox 2} respectively). All the users will use an Internet browser to access the tools in the sandbox and all the interactions are mediated by the broker.

The \emph{Sandbox Broker} has access to the catalogues of different models, tools, and pre-configured OSs that are available, so an end user can simply pick a valid combination to request a sandbox. In addition to those catalogues, the \emph{Sandbox Broker} keeps user information, such as the user's models (private copies of the model in the catalogue, which may have been modified by the user while using the sandbox) and the sandboxes the user created. This information is important to allow the creation of new sandboxes.

The operation of user requests and the sandboxing logic is provided by the \emph{Sandboxing Kernel}, which is a component that interacts with the system \emph{Hypervisor} to launch the different constituents of a sandbox, namely:
\begin{itemize}
	\item \textit{NFS} - Network File System providing storage in the form of shared folders where model files and tool outputs are placed.
	\item \textit{VLANS} - Virtual networks restricting the communications of the VMs inside a sandbox to the set of VMs composing it and those only.
	\item \textit{V.OS} - Virtual machines running the OSs supporting a tool, a remote desktop protocol to provide the clients access to the tool display, and other monitor and interoperability tools to operate the VM inside the Kernel.
	\item \textit{Tools} - The tools running a model or a multi-model.
	\item \textit{Models} - A mathematical/formal description of a component.
\end{itemize}

The operation relies on a database of metadata about the different sandboxes (the \emph{Sandboxes Metadata} component in the figure). This component stores and keeps track of the sandboxes' states (running, suspended, \dots) and user ownership of the resources. It is worth to highlight is that the Kernel has direct network connections to the Sandboxes' VLANs.

\subsection{Security}

To ensure the data privacy of models and analysis results produced in the HUBCAP experiments, we envisage a security middleware to enhance the HUBCAP Platform with  the due security and protection mechanisms.
The sandbox design itself should ease security auditing and assurance. For instance, our current proposal follows a trusted ``kernel architecture'' where the sandboxes manager/broker launches sandboxes and mediates its interactions. An example of such is the multi-process architecture design of Google Chrome, where each tab untrusted code (not necessarily malicious, but not possible to be assumed secure) runs in a sandbox environment and accesses the system resources through a trusted broker ensuring independence of the different tabs. Thus it is possible to secure the platform by a general verification approach \cite{Jomaa&18}.

Moreover, the components of the sandbox kernel are open source, e.g.\
Hypervisor, NFS server, and the security will be based on Data-Service
Sovereignty principles, in order to enhance trust among beneficiaries of the
HUBCAP Platform (exchanging data and services), but also intended to provide
protection mechanisms to prevent the infiltration of malware into the
collaboration platform by applying known malware detection techniques, for
instance \cite{Macedo&13}, which will systematically check the collaboration
platform FMUs for malicious behaviour.  Furthermore, secure isolation
\cite{Suciu&18} and security information and event management (SIEM) can ensure
that aggregated data and log records can be automatically analysed giving a
clear picture of what is happening on the platform.

\section{\uppercase{Catalogues of Models and MBD services}}\label{sec:models}

The HUBCAP Platform will provide access to assets
including models and tools for MBD. The ana\-ly\-sis capabilities of the
tools will be available as services to be tested in a sandbox. The MBD
services will include support for modelling of CPS with components,
contracts, and equations, for analysis based on simulation, model
checking, model-based safety analysis, for synthesis of HW/SW
deployments, fault detection and recovery, planning, and for many more
functionalities. We anticipate that the platform's user community will integrate
more tools and models over time.

Models and services will be presented to the user in
catalogues, where the users will choose the tool, the kind of analysis
they want to try, and existing models associated to it to exemplify
the usage. The HUBCAP Platform will create a dedicated sandbox
with the tool already installed and the desired models ready to be used. The
user will follow the instructions to perform the analysis on the
chosen model. The users will be able to write their own models and test the
capabilities of the tool. If needed, the users will be able to get support by the
tool experts via the collaboration services of the Platform.

\subsection{Model Catalogue}

We expect that users will supply adaptable and generic models
to assist newcomers to specific modelling tools and tool combinations. The Model Catalogue
will support access to these. Initially, we would expect to include models from standards and tutorials. For example, we expect initially to include all the public models that are available
for the INTO-CPS tool chain \cite{Mansfield&17} that can be imported and used directly
inside the tools. Other example would include the wheel braking system architecture and
the related fault trees described in the AIR6110 standard \cite{AIR6110} and
modelled in \cite{BozzanoCPJKPRT15}.

\subsection{MBD Services}\label{sec:mdbservices}

In the following we provide examples of tools and services that may be realised
using the HUBCAP Platform. To provide further detail in the conceptualisation
of the sandbox service, we organise the examples by the kind of
sandbox defined in Section~\ref{sec:tsf}.


\paragraph{Example of a Demo Sandbox.} Controllab Products is a tool provider
SME
Among other products that may be hosted in other kinds of sandboxes,
they provide pay-as-you-go access to virtual reality 3D animations of
model's simulations. The animations require high-end computer specifications
and specialised VR hardware. In such cases hosting the experiments on a remote
sandbox hosted on the HUBCAP Platform is neither feasible nor desired. Thus
the sandbox may provide a simple web application with highlights of the full
content and provide the pointers to the replication of the experiment
environment on an adequate hardware setup.

\paragraph{The Container Example.} BEIA Consult is an SME that has
developed models and a tool applied in energy efficiency of smart buildings \cite{Suciu&19}.
The application is containerised using an
open source platform, i.e.\ the tool is stored in containers/software packages
that are run by a platform/container orchestration software such as Docker. Therefore, a
sandbox with a single V.OS
hosting the container orchestration software is
available to execute the tool and models.

\paragraph{The Cloud Example.} In this case a combination of Linux-based
tools and  Windows-only tools can be used to edit/display model details and
generate Functional Mock-Units (FMUs). The FMUs may be then used to perform a
co-simulation, the simulation of the behaviour of the joint executions of the
different models.  The co-simulation experiment may run in a third virtual
machine hosting a containerised/cloud version of the INTO-CPS co-simulation
tool \cite{Macedo&20}. Although the INTO-CPS tool could be run in a single V.OS
to perform a simulation of diverse models, the access to the edition and
visualization of the models requires other tools, which is a feature that makes
the INTO-CPS tool a good fit for the Cloud Sandbox service level.

\section{\uppercase{Open Calls}}\label{sec:opencalls}

To encourage the population of the platform, and to evaluate its use as an aid
to innovation in CPS design, HUBCAP will run a series of funded Open Calls.
These will provide financial and
technical support for SMEs to join the HUBCAP ecosystem and to experiment in
highly innovative, cross-border experiments. There are three series of calls,
each with different purposes:

\paragraph{PULL Calls}
``PULL'' calls encourage the population of the platform by model and tool suppliers. SMEs may request awards (up to \euro{}1,000) to aid in covering the costs of integration into the platform, including participation in a one-day workshop and
3 -- 4 days overall effort.  There will be five such calls and we expect to sponsor 200 projects.

\paragraph{EXPERIMENT Calls.}
``EXPERIMENT'' calls will support consortia of typical two SMEs experiment with the adoption of MBD for CPSs using assets and services from the platform, in particular from SMEs with less prior digital experience. Consortia may bid for \euro{}30,000 - \euro{}75,000
for projects of 4 to 6 months duration.  There will be two \textbf{EXPERIMENT}
calls and we expect to fund 20 to 30 projects.

\paragraph{INNOVATE Call.}
One \textbf{INNOVATE} call will up to \euro{}200,000 for consortia of SMEs to
deploy new products and demonstrate highly innovative collaborations using the
HUBCAP Platform. Funding supports a project of 12 months' duration.  There will
be one such call and we expect to grant 10 projects.


According to \cite{Prato&15}, 63\% of high potential
innovations arise within projects
in their final stages and 41\% of all
organisations behind these are SMEs.  Therefore, as it is expected that new
innovations will emerge towards the third year from the project third-party
experiments, the \textbf{EXPERIMENT} and \textbf{INNOVATE} calls are supplied
with a larger budget and are open later in the project timeline. Conversely,
the \textbf{PULL} call is open from early in the project (with five regular
deadlines) and has a smaller budget dedicated to workshops helping partners get
their assets into the collaboration platform.

\section{\uppercase{Concluding Remarks and Future Work}}\label{sec:conclude}

The HUBCAP Platform is still under development, but we expect that the first public version will become available in late 2020. It is intended to form a shared resource for an ecosystem in which diverse organisations will supply models and tools to encourage and ease the evaluation and adoption of MBD approaches for CPSs.

Our hope is that the HUBCAP ecosystem supported by this platform might
encourage development of MBD through servitisation. In
the future, users and tool suppliers will explore, share, and buy CPS assets
(models, tools, services, training) from across the ecosystem through a
``test-before-invest'' sandbox and -- at least in some cases -- integrated ``pay-as-you-go'' charging.

We anticipate that, in the course of populating the collaboration platform, we will run into limitations in the capabilities of both tools and the sandbox architecture. It is envisaged that there will be challenges in regards to both licenses for OSs as well as tools with special needs for particular hardware support (e.g., graphics cards) that may not easily be supported in a sandbox context.

We hope that the HUBCAP Platform will be extended in several directions enabling true collaboration between different participating organisations, alongside and as a result of the open calls. We envisage that the HUBCAP Platform may be conveniently hosted at standard cloud operators as well as on servers at large companies with many suppliers such that they can be in full control of the development of the collaboration around larger CPSs such as automobiles and airplanes.

\subsubsection*{Acknowledgements.}
The work presented here is partially supported by the HUBCAP Innovation Action funded by the European Commission's Horizon 2020 Programme under Grant Agreement 872698. We would also like to thank Claudio Gomes for input on drafts of this paper.

\bibliographystyle{apalike}
{\small
\bibliography{../bib/hubcap}

\newcommand{\noop}[1]{}
\begin{thebibliography}{}

\bibitem[Beckers et~al., 2007]{Beckers&07}
Beckers, J., Heemels, M., Bukkems, B., and Muller, G. (2007).
\newblock Effective industrial modeling: The example of happy flow.
\newblock In Heemels, M. and Muller, G., editors, {\em Boderc: Model-based
  design of high-tech systems}, chapter~6, pages 77--88. Embedded Systems
  Institute, Den Dolech 2, Eindhoven, The Netherlands, second edition.

\bibitem[Blochwitz, 2014]{FMIStandard2.0}
Blochwitz, T. (2014).
\newblock {Functional Mock-up Interface for Model Exchange and Co-Simulation}.
\newblock \url{https://www.fmi-standard.org/downloads}.

\bibitem[Bozzano et~al., 2015]{BozzanoCPJKPRT15}
Bozzano, M., Cimatti, A., Pires, A.~F., Jones, D., Kimberly, G., Petri, T.,
  Robinson, R., and Tonetta, S. (2015).
\newblock {Formal Design and Safety Analysis of {AIR6110} Wheel Brake System}.
\newblock In {\em CAV}, pages 518--535.

\bibitem[Cook, 2008]{Cook08}
Cook, N. (2008).
\newblock {\em Enterprise 2.0: How social software will change the future of
  work}.
\newblock Gower Publishing Limited.

\bibitem[De~Prato et~al., 2015]{Prato&15}
De~Prato, G., Nepelski, D., Piroli, G., et~al. (2015).
\newblock {Innovation Radar: Identifying Innovations and Innovators with High
  Potential in ICT FP7, CIP \& H2020 Projects}.
\newblock {\em Science and Policy Report, Joint Research Centre}.

\bibitem[Fritzson, 2015]{Fritzson15}
Fritzson, P. (2015).
\newblock {\em Principles of Object-Oriented Modeling and Simulation with
  Modelica 3.3: A Cyber-Physical Approach}.
\newblock IEEE Press. Wiley, 2 edition.

\bibitem[Gomes et~al., 2018]{Gomes&18}
Gomes, C., Thule, C., Broman, D., Larsen, P.~G., and Vangheluwe, H. (2018).
\newblock {Co-simulation: a Survey}.
\newblock {\em ACM Comput. Surv.}, 51(3):49:1--49:33.

\bibitem[Jomaa et~al., 2018]{Jomaa&18}
Jomaa, N., Nowak, D., Grimaud, G., and Hym, S. (2018).
\newblock Formal proof of dynamic memory isolation based on mmu.
\newblock {\em Science of Computer Programming}, 162:76--92.

\bibitem[Kainz et~al., 2019]{Kainz&19}
Kainz, O., Jakab, F., Michalko, M., Hud{\'a}k, M., and Petija, R. (2019).
\newblock Enhanced approaches to automated monitoring environmental quality in
  non-isolated thermodynamic system.
\newblock {\em IFAC-PapersOnLine}, 52(27):365--376.

\bibitem[Larsen et~al., 2016]{Larsen&16a}
Larsen, P.~G., Fitzgerald, J., Woodcock, J., Fritzson, P., Brauer, J., Kleijn,
  C., Lecomte, T., Pfeil, M., Green, O., Basagiannis, S., and Sadovykh, A.
  (2016).
\newblock {Integrated Tool Chain for Model-based Design of Cyber-Physical
  Systems: The INTO-CPS Project}.
\newblock In {\em CPS Data Workshop}, Vienna, Austria.

\bibitem[Macedo et~al., 2020]{Macedo&20}
Macedo, H.~D., Sanjari, A., Villadsen, K., Thule, C., and Larsen, P.~G. (2020).
\newblock {Introducing Angular Tests and Upgrades to the INTO-CPS Application}.
\newblock In {\em Submitted for publication}.

\bibitem[Macedo and Touili, 2013]{Macedo&13}
Macedo, H.~D. and Touili, T. (2013).
\newblock Mining malware specifications through static reachability analysis.
\newblock In {\em European Symposium on Research in Computer Security}, pages
  517--535, Berlin, Heidelberg. Springer, Springer Berlin Heidelberg.

\bibitem[Mansfield et~al., 2017]{Mansfield&17}
Mansfield, M., Gamble, C., Pierce, K., Fitzgerald, J., Foster, S., Thule, C.,
  and Nilsson, R. (2017).
\newblock Examples compendium 3.
\newblock Technical report, The {INTO-CPS} Project.

\bibitem[SAE, 2011]{AIR6110}
SAE (2011).
\newblock {AIR 6110, Contiguous Aircraft/System Development Process Example}.

\bibitem[Suciu et~al., 2018]{Suciu&18}
Suciu, G., Istrate, C., Petrache, A., Schlachet, D., and Buteau, T. (2018).
\newblock On demand secure isolation using security models for different system
  management platforms.
\newblock In {\em Advanced Topics in Optoelectronics, Microelectronics, and
  Nanotechnologies IX}, volume 10977, page 109770R. International Society for
  Optics and Photonics.

\bibitem[Suciu et~al., 2019]{Suciu&19}
Suciu, G., Necula, L., Iosu, R., Usurelu, T., and Ceaparu, M. (2019).
\newblock Iot and cloud-based energy monitoring and simulation platform.
\newblock In {\em 2019 11th International Symposium on Advanced Topics in
  Electrical Engineering (ATEE)}, pages 1--4. IEEE.

\bibitem[Sztipanovits et~al., 2015]{Sztipanovits&15}
Sztipanovits, J., Bapty, T., Neema, S., Koutsoukos, X., and Jackson, E. (2015).
\newblock {Design Tool Chain for Cyber-physical Systems: Lessons Learned}.
\newblock In {\em Proceedings of the 52nd Annual Design Automation Conference},
  DAC '15, pages 81:1--81:6, New York, NY, USA. ACM.

\bibitem[Van~der Auweraer et~al., 2013]{VanderAuweraer&13}
Van~der Auweraer, H., Anthonis, J., De~Bruyne, S., and Leuridan, J. (2013).
\newblock Virtual engineering at work: the challenges for designing mechatronic
  products.
\newblock {\em Engineering with Computers}, 29(3):389--408.

\end{thebibliography}

}

\end{document}